\documentclass[12pt]{article}
\usepackage{changebar}
\usepackage{epsfig}
\usepackage{amsfonts}
\usepackage{amsmath}
\usepackage{amssymb}
\usepackage{amscd}
\usepackage{afterpage}
\usepackage{subeqnarray}
\usepackage{float}
\usepackage{amsbsy}
\usepackage{feynmf}

\begin{document}

\def\bfm#1{{\mbox{\boldmath $#1$}}}
\def\intt#1{\int \frac{d^3#1}{(2\pi)^3}}
\def\intq#1{\int \frac{d^4#1}{(2\pi)^4}}
\def\intx#1{\int \frac{d^3#1}{(2\pi)^3}\frac{1}{2#1_0}}
\def\P{{\cal P}}
\def\Q{{\cal Q}}
\def\sumint{{\sum\hskip-15pt\int}}
\def\Gh{{G_{\rm h}}}
\def\Gp{{G_{\rm p}}}
\def\gh{{{\cal G}_{\rm h}^{\rm s.m.}}}
\def\Mh{{{\cal M}_{\rm h}}}
\def\Mp{{{\cal M}_{\rm p}}}
\def\Tr{{\rm Tr}}

\author{R. Cenni${}^\dagger$ and G. Vagradov${}^\dagger$
\\
${}^\dagger$Istituto Nazionale di Fisica Nucleare -- sez. di Genova\\
Dipartimento di Fisica dell'Universit\`a di Genova\\
Via Dodecaneso 33 -- 16146 -- Genova -- Italy\\
${}^\ddagger$
Institute for Nuclear Research --
117312 Moscow Russia
}

\title{On the Relativistic Description of the Nucleus}

\maketitle

\begin{abstract}
  We present a formalism able to generalise to a relativistically 
  covariant
  scheme the standard nuclear shell model. We show that, using some generalised
  nuclear Green's functions and their Lehmann representation we can define
  the relativistic equivalent of the non relativistic 
  single particle wave function (not loosing, however, the physical 
  contribution of other degrees of freedom, like mesons and antinucleons).
  It is shown that the mass operator associated to the nuclear Green's function
  can be approximated with the equivalent of a shell-model potential and
  that the corresponding ``single particle wave functions'' can be 
  easily derived in a specified frame of reference and then boosted to 
  any other system, thus fully restoring the Lorentz covariance.
\end{abstract}

\section{Introduction}

The difficulties one meets in building a theory for relativistic bound 
system with finite number of particles are well known. Up to now,
in spite of many efforts in this field (see \cite{SeWa-86,CeSh-86-B}
for a comprehensive review, but also \cite{Ra-01}
to get an example of the present approach to the relativistic shell model), 
to reconcile relativity,
translational invariance and shell model seems to be a very hard task. 
Moreover, even the connection between exclusive and inclusive
processes is non-trivial for two order of 
problems.

By one side in fact  in a relativistic 
framework it is impossible to fully disentangle the nucleonic and nuclear 
dynamics \cite{Al-al-88} 
(in a few word the nucleon form factors do not  factorize)
even in the simple scheme of the Plane Wave Impulse Approximation (PWIA), 
because  the separation between longitudinal and transverse motion
is a frame-dependent concept and the Fermi motion of the nuclei prevents
a full separation of them in a nuclear context. Moreover even the concept
of Coulomb sum rule as it is usually interpreted looses its meaning and can be
regained only at the prize of introducing a suitable renormalization
factor (that, fortunately, turns out to be largely model 
independent)\cite{CeDoMo-97}.

On the other side, when going beyond PWIA multiple
counting of diagrams occurs \cite{CeVa-95}, 
with the consequence that the integral of
the differential cross sections for $(e,e^\prime p)$ or $(e,e^\prime n)$ 
reactions  no longer coincide with the inclusive cross section (this because
of the existence of channels where, for instance, another nucleon is emitted 
but not revealed).

Thus with the usual many-body expansion it is very difficult
to connect forward scattering amplitude with the total cross section.
This suggested us to extend the idea of Green's functions (of any kind)
by allowing situations where the kinematics of the initial and final nucleus 
can be different. This will be suitable to directly study the elastic processes
in a fully Poincar\'e invariant way, but natural extensions could also be
obtained (and we plan do pursue this line in the future) by choosing
different initial and final states.

For the moment we limit ourselves 
to the problem of two interacting particles, namely a nucleus $\pm$ a nucleon 
or (if case) an elementary particle (nucleon)
$\pm$ a quark. This job enables us to account for recoil effects in high 
energy nuclear reactions and in quark physics (excitations of nucleon, meson).

We first begin in section \ref{sec:0} with a short review of what it happens in
the non-relativistic frame, in order to provide a layout of the matter
we would like to generalise, and also in order to make easier the understanding
of the origin of some problem we are concerned with, i.e., if they arise
from the many-body theory or from the relativity.

Next, in Section \ref{sec:1} we consider the one particle (or hole) problem in 
presence
of a nucleus (to be more specific we consider nuclei with $A$ nucleons $\pm$ 
one nucleon).
We use a formalism similar to the one of the Green's function formalism. 
Due to nuclear recoil the equations for hole and particle become different 
unlike the  infinite systems or  models where recoil is neglected.

In Section \ref{sec:4}
 we introduce a model with particle (hole)-nucleus interaction,
conceptually similar to the shell model. The equation of motion
in such approach looks like the equation for a particle embedded in a
 mean field, but
the equation is relativistically covariant (of course with suitable choice of
the interaction). The crucial point here is that we need to introduce the 
"shell" ground state of the system.

Section \ref{sec:5x} provides a possible  perturbative scheme to go beyond the 
mean field level, and sec. \ref{sec:6} presents a simple case  where the 
``relativistic shell model'' can be easily solved, thus coming in contact 
with the real world, not pursuing abstract and inapplicable 
theoretical formalisms.

\section{The non-relativistic single particle motion in nuclei}
\label{sec:0}

The purpose of this section is twofold: by one side to remind the reader the
general scheme of some old models for the propagation of a particle (or
hole) inside a nucleus, that pertain to the history of nuclear physics, but
that  (partially) could be scarcely useful today in practical calculations;
on the other side we wish to remind those topics that can provide a 
guideline for our generalisation to a relativistic finite nucleus  
and to ``non-diagonal'' Green's functions (this topic well be clarified
in the next section).

If we limit ourselves to the propagation of a particle (or hole) inside a 
nucleus the most natural framework to begin with is certainly the
Feschbach's approach to the optical potential\cite{Fe-58,Fe-62}. 
Let us repeat once more
its main topics (or, better, the ones we need in the 
following). The general idea is that, if a Hamiltonian is defined in a
Hilbert space $\cal H$, then we can project Schr\"odinger equation into
a subspace ${\cal H}^\prime\subset{\cal H}$ the price to be payed being an
energy dependence in the effective potential. As everybody knows, if $\P$
is a projection operator
$$\P:{\cal H}\longrightarrow {\cal H}^\prime$$
and $\Q=I-\P$ then the Schr\"odinger equation in the space ${\cal H}^\prime$
reads
\begin{equation}
  \label{eq-1:1}
  {\cal H}_{\rm opt}(E)| \Psi>=
  \left[\P H\P+\P H\Q \frac{1}{E-\Q H \Q+i\alpha}\Q H \P\right]\P| \Psi>=
  E\P| \Psi>
\end{equation}
The conceptual points we want to remark are the following:
\begin{itemize}
\item the optical potential can be defined in this way and its   most relevant
structures can be derived, but eq. (\ref{eq-1:1}) can by no means be used to
evaluate it.

\item Eq. (\ref{eq-1:1}) is quite general: according to the definition of 
$\P$ it can be adapted to a variety of problems: we shall consider in the
following the particle and hole propagation in a nucleus but we 
shall also remind
its application to $(e,e^\prime p)$ reactions in impulse approximation.
\item Eq. (\ref{eq-1:1}) specifically imposes causality at each time.
We shall see that this ``microscopic'' causality is the first reason that
inhibits a microscopical calculation of the optical potential.

\item 
The optical potential displays an imaginary part, but since (\ref{eq-1:1})
is derived from a true Schr\"odinger equation in a bigger space, the 
eigenvalues are necessarily real. This applies of course to the discrete
ones, i.e., to stable nuclear states. In order to conserve a Lehmann
representation with real eigenvalues, the usual way out is that of
discretizing the whole system by means of a box normalisation.
\end{itemize}

The next requirement is the definition of $\P$, i.e., of the physical problem
we will concerned with. In the archetypal case, namely the elastic scattering
of protons off nuclei $\P$ reads
\begin{equation}
  \label{eq-1:2}
  \P=\int d^3r\,d^3r^\prime \rho({\bf r},{\bf r^\prime})
\psi^\dagger({\bf r})|\Phi_A^0><\Phi_A^0|
\psi({\bf r}^\prime)
\end{equation}
where $|\Phi_A^0>$ is the ground state of a nucleons with A nucleons,
$\psi$ and $\psi^\dagger$ are the non-relativistic destruction and creation 
operator of a nucleon in the point ${\bf r}$ (spin and isospin will be 
neglected throughout this paper) and $\rho$ is fixed by the requirement
$\P=\P^2$, that implies symbolically
\begin{equation}
  \label{eq-1:3}
  \rho={1\over I-n}
\end{equation}
where 
\begin{equation}
  \label{eq-1:4}
  n({\bf r},{\bf r^\prime})=<\Phi_A^0|\psi^\dagger({\bf r})\psi({\bf r}^\prime
)|\Phi_A^0>\;.
\end{equation}

The space of the  solutions of the eigenvalue equation (\ref{eq-1:1}) is of
course isomorphic to $L^2({\mathbb R}^3)$. We can define an orthonormal basis
in $L^2({\mathbb R}^3)$ by defining 
\begin{equation}
  \label{eq-1:5}
  |{\bf r})\equiv \int d^3 r^\prime \left\{I-n\right\}^{-{1\over 2}}
({\bf r},{\bf r}^\prime)\psi^\dagger ({\bf r}^\prime)|\Phi_A^0>\;.
\end{equation}
We also assume that this basis is complete. The eigenvalue equation now reads
\begin{equation}
  \label{eq-1:6}
  \int d^3 r^\prime ({\bf r}|{\cal H}_{\rm opt}(E)|{\bf r}^\prime)
\varphi_n({\bf r}^\prime)=E\varphi_n({\bf r})
\end{equation}
and we are in position to connect the solutions of (\ref{eq-1:6}) with the
true eigenstates of the $A+1$ system: let $|\Phi_{A+1}^n>$ a solution of the 
complete Schr\"odinger equation in the space of $A+1$ particles. 
Some simple algebra enable us to write the solutions of (\ref{eq-1:6})
in the form
\begin{equation}
  \label{eq-1:7}
  \varphi_n({\bf r})=\int d^3 r^\prime \left\{I-n\right\}^{-{1\over 2}}
({\bf r},{\bf r}^\prime)<\Phi_A^0|\psi({\bf r}^\prime)|\Phi_{A+1}^n>
\end{equation}
the eigenvalue being of course $E_n$.
For future reference let us define the function
\begin{equation}
  \label{eq-1:8}
  \psi_n({\bf r})=<\Phi_{A+1}^n|\psi^\dagger({\bf r})|\Phi_A^0>
\end{equation}
We have shown above that up to a rescaling $\psi_n$ is solution of the
eigenvalue equation for the ``optical'' Hamiltonian.
It also follows that  $\psi_n$ is, by itself, eigenstate of the (unsymmetric)
operator
$$\sqrt{1-n} {\cal H}_{\rm opt} \frac{1}{\sqrt{1-n}}~.$$

Very much in the same way we can handle other situations. In particular
we shall be concerned with hole propagation inside a nucleus. Thus we define
another projection operator, namely
\begin{equation}
  \label{eq-1:9}
  \tilde\P=\int d^3r d^3 r^\prime \psi({\bf r})|\Phi_A^0>
n^{-1}({\bf r},{\bf r}^\prime)
<\Phi_A^0|\psi^\dagger({\bf r}^\prime)
\end{equation}
and all the above formalism follows up provided we do the substitutions
$$I-n\longrightarrow n$$
and
\begin{equation}
  \label{eq-1:10}
  \psi_n({\bf r})\longrightarrow\phi_n({\bf r})=<\Phi_{A-1}^n|
\psi ({\bf r})|\Phi_A^0>~.
\end{equation}

The next step, in the early times of the optical potential, was the connection
with the single particle Green's function of the system, defined, as usual,
as
\begin{equation}
  \label{eq-1:11}
  G(x,x^\prime)=\frac{<\Phi_A^0|{\cal T}\left\{\psi(x),\psi^\dagger(x^\prime)
 \right\}|\Phi_A^0>}{<\Phi_A^0|\Phi_A^0>}\;.
\end{equation}
It was shown by Bell and Squires\cite{BeSq-59}
that the self-energy (or mass operator)
 can be interpreted as an optical potential (not coincident
however with the one introduced by Feshbach and discussed above).
As is well known  $G$ can be separated into a retarded and an advanced
part having the following Lehmann representation:
\begin{eqnarray}
  \label{eq-1:12}
  G&=&G^++G^-\\
\label{eq-1:13}
  G^+({\bf r},{\bf r}^\prime;\omega)&=&
\sum_n{\psi_n({\bf r})\psi^*({\bf r}^\prime)
 \over \omega-E_n^{A+1}+i\alpha}\\
\label{eq-1:14}
 G^-({\bf r},{\bf r}^\prime;\omega)&=&\sum_n
{\phi_n({\bf r})\phi^*({\bf r}^\prime)
 \over \omega-E_n^{A-1}-i\alpha}
\end{eqnarray}
where the functions $\psi_n$ and $\phi_n$
are  those discussed above. Note that the
functions $\psi_n$ and $\phi_n$ are in some way connected with the shell model:
in fact the index $n$ runs over all the possible nuclear eigenstates,
but grouping together some levels and constructing in this way the 
single-particle levels and discarding those with a too small strength one
is lead back to the shell model. This however implies the breaking of the
translational invariance, since the latter would strictly imply the
functional dependence
$$\phi_n,\psi_n\sim e^{i{\mathbf p\cdot r}}~.$$
How to recover the translational invariance and at the same time to leave
sufficient room t2o introduce the analogous of the ``shell-model wave 
functions'' will be the task pursued in the following. 

The previous discussion enables us to write down (but by no means to solve or
to approximate) the inverse of $G^\pm$. We can write indeed
\begin{equation}
  \label{eq-1:15}
  \left[G^\pm\right]^{-1}({\bf r},{\bf r}^\prime;\omega)=\omega-T
-{\cal M}^\pm({\bf r},{\bf r}^\prime;\omega)
\end{equation}
where ${\cal M}^\pm$ could be called the mass operator for particles or holes,
with the property
\begin{equation}
  \label{eq-1:16}
  G^+\psi^*_n({\bf r})=0\qquad\qquad\qquad G^-\phi^*_n({\bf r})=0
\end{equation}
for $\omega=E_n^{A\pm1}$. The discussion above shows that ultimately
\begin{equation}
  \label{eq-1:17}
  G^\pm=\sqrt{1-n} {\cal H}_{\rm opt} \frac{1}{\sqrt{1-n}}
\end{equation}
provided the particle or hole projection operator is used in the rhs.
In this way we have indirectly defined the mass operators ${\cal M}^\pm$;
it must be reminded however that the whole Green's function obeys the
Dyson's equation
\begin{equation}
  \label{eq-1:18}
  G=G^0+G^0 {\cal M}G
\end{equation}
where the mass operator (or self-energy) can be derived from a perturbative
expansion, but it is {\em not} the sum of the two mass operators defined 
separately for particles and holes, and they can in no way be derived from
any perturbative scheme.

Before ending this section we would also remind that the same formalism have 
been employed in studying $(e,e^\prime p)$ reactions in the frame of the 
Distorted Wave Impulse Approximation (DWIA). There it is 
convenient to define many projection operators, any of them pertaining to a 
residual nucleus left in an excited state plus an outgoing nucleon.
This again is formally correct but by no means one can be able to explicitly
write down the (almost) infinite set of different optical potentials.
Thus one ultimately ends up with assuming the same  optical potential
for the outgoing nucleon independently of the state the residual nucleus is
left in. As a non-trivial consequence the differences between longitudinal
and transverse channels are lost (see \cite{CeCiSa-89} to recover it). 
Again here the most natural 
treatment of the problem goes through the introduction of a particle-hole
Green's function having the form (we follow the standard notations)
\begin{equation}
  \label{eq-1:18.1}
  \Pi^{\mu\nu}(x,x^\prime)=\frac{<\Phi_A^0|{\cal T}\left\{
j_\mu(x),j_\nu(x^\prime)
 \right\}|\Phi_A^0>}{<\Phi_A^0|\Phi_A^0>}\;,
\end{equation} 
where $j_\mu$ is the electro-magnetic current. 
The differences (we could better say the incompatibility) between this approach
and the DWIA has already been shown by the authors of this paper in ref. 
\cite{CeVa-95}.

\section{The generalised one-body Green's function }
\label{sec:1}

In a relativistic approach, with the aim of pursuing the analogy
with the description of the non-relativistic single-particle or single-hole 
motion discussed above, and moreover in order to avoid the disease of multiple
counting of diagrams as outlined in \cite{CeVa-95}, we consider
a bound system of fermionic and bosonic fields with 
finite baryonic number and in the ground state in its frame of reference
but assuming in general  non-zero different total momenta
for incoming and outgoing states.

This approach will be applied here to nuclear physics, but the study of the
 quarks dynamics in a nucleon could also be an affordable task. Moreover
in both cases the accounting of the recoil effects is allowed.

Let us first of all introduce the incoming and outgoing nuclear bound state
$|p>$ and $p^\prime>$ for a nucleus of mass $M$ and initial and final 3-momenta
${\bf p}$ and ${\bf p^\prime}$. We can of course introduce initial and final
4-momenta by putting 
$p_0=\sqrt{{\bf p}^2+M^2}$ and
$p_0^\prime=\sqrt{{{\bf p}^\prime}^2+M^2}$.
The normalisation reads
\begin{equation}
  \label{eq0}
  <p^\prime|p>=(2\pi)^32p_0\delta^3({\bf p}-{\bf p}^\prime)\;.
\end{equation}

We now define a generalised single particle 
Green's function as
\begin{equation}
  \label{eq0.1}
  G_{pp^\prime}(y,y^\prime)=\frac{1}{2\sqrt{p_0p_0'}}
  < p^\prime|{\cal T}\left\{\psi(y),\overline
\psi(y^\prime)\right\}|p>\;.
\end{equation}
We have already observed in the previous section that in a 2-points Green's
function the translational invariance 
will rigidly constrain the analytical form of the functions
$\phi_n$ and $\psi_n$ thus forbidding its interpretation (within some
approximation schemes) as single-particle wave functions.
The above choice of writing a generalised single particle 
Green's function, with, actually, one more argument, relaxes the above 
constraint and will turn out to be the key issue in constructing
the relativistic generalisation of the shell model without violating
the Poincar\'e invariance.

The particular case we are considering deserves a comment about the realization
of the linked cluster theorem.
To understand it we could interpret $ G_{pp^\prime}(y,y^\prime)$
as a limiting case of a two-particle Green's function. Imagine that
$\Psi({\bf p},t)$ is the destruction operator of the nucleus in its ground 
state with total momentum ${\bf p}$ and $|0>$ the physical vacuum. Then
$ G_{pp^\prime}(y,y^\prime)$ can be regarded (up to normalising factors)
 as the following limit:
$$\lim_{t\to-\infty}\lim_{t^\prime\to+\infty}{<0|
 {\cal T}\left\{\psi(y),\overline\psi(y^\prime),\Psi^\dagger({\bf p},
t),\Psi({\bf p}^\prime,t^\prime)\right\}|0>\over <0|0>}$$
and the denominator $<0|0>$ is the tool that ensures the cancellation 
of the disconnected diagrams in the two (composite)  
particle Green's function. Of
course this property is preserved through the limiting process and 
consequently $ G_{pp^\prime}(y,y^\prime)$ (where we neglect the denominator
 $<0|0>$ throughout this paper) has always to be intended as constructed
by linked diagrams only.

Now we want to represent the function $ G_{pp^\prime}(y,y^\prime)$ in
Lehmann representation. First however we need some kinematical considerations.

\begin{figure}[h]
  \begin{center}
    \centerline{
      \epsfig{file=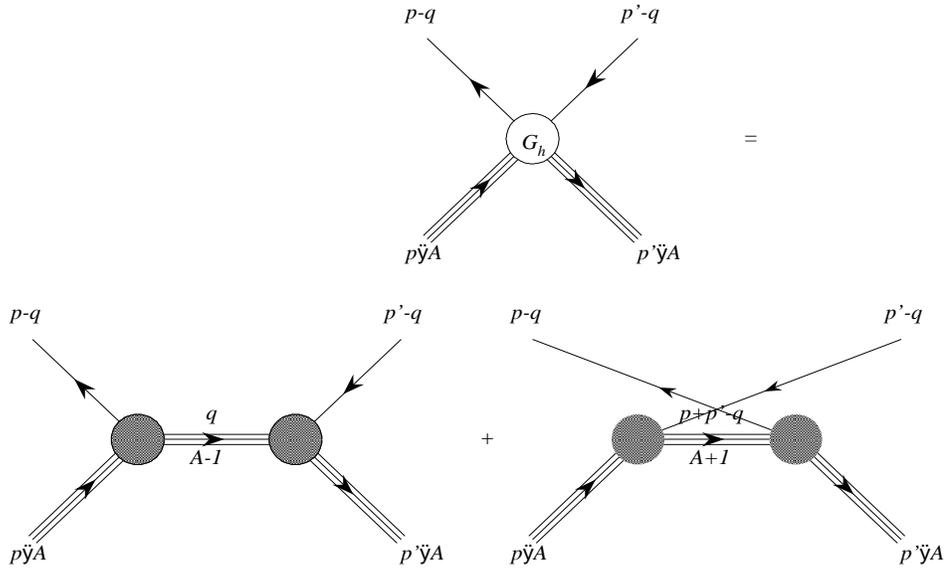,height=8cm,width=14cm}
      }
    \caption{The kinematics for the ``hole'' propagation}
    \label{fig:10}
  \end{center}
\end{figure}

If $\hat p$ is the 4-momentum operator, i.e., $\hat p=(\hat p_0=\hat H,\
\hat {\bf p})$,
$$\hat p_\alpha |p>=p_\alpha |p>$$
then we know that
\begin{equation}
  \label{eq0.5}
  \psi(y)=e^{i\hat p\cdot y}\psi e^{-i\hat p\cdot y}
\end{equation}
(with $\psi\equiv\psi(y=0)$). With these definitions we can write the Fourier
transform of  $ G_{pp^\prime}(y,y^\prime)$.
Here however an ambiguity arises, since our $G$ is ultimately, 
as quoted above, 
a two-particle Green's function, we can choose  in Fourier transform two 
different kinematics, one tailored for the propagation of a particle 
and one for a hole.
For the ``hole'' channel, whose kinematics is depicted in fig. \ref{fig:10}
we define
\begin{eqnarray}
  \label{eq0.5.1}
  \Gh(p,p^\prime;q,q^\prime)&=&\int d^4y\,d^4y^\prime
e^{i(p-q)y-i(,p^\prime-q)y^\prime}G_{pp^\prime}(y,y^\prime)\nonumber\\
&=&(2\pi)^4\delta^4(q-q^\prime)\Gh(p,p^\prime;q)
\end{eqnarray}
Let us remark that all the formalism here is covariant, and in order remind
this property we indicate in $G_h$ a dependence upon the 4-vectors $p$,
$p^\prime$ and $q$. Strictly speaking the mass shell condition on $p$ and
$p^\prime$ would entail a dependence upon the 3-vector part only.

The Lehmann representation of $\Gh$ is easily written down as
\begin{equation}
  \label{eq1}
  \Gh(p,p^\prime;q)=
  \frac{(2\pi)^3}{2\sqrt{p_0p'_0}}
  <p^\prime|\overline\psi{\delta(\hat{\bf p}-{\bf q})
    \over H-q_0-i\alpha}\psi+\psi{\delta(\hat{\bf p}-
    {\bf q}')\over q_0^\prime-H+i\alpha}\overline\psi|p>
\end{equation}
where
\begin{equation}
  \label{eq:25n}
  q'=p+p'-q~,
\end{equation}
that reflects the kinematics of fig. \ref{fig:10}. There 
the intermediate lines denote the sum over all the eigenstates of the system
with baryonic number $A-1$ (first term) or $A+1$ 
(second term in the rhs of fig. \ref{fig:10}) having total 4-momentum $q$. 
The former states are 
characterised by
\begin{equation}
\label{eq:1o}
H|q_\lambda>=q_{\lambda 0}|q_\lambda>\;.
\end{equation}
where $\lambda$ is an index running over all the $A-1$ states, their mass
being $M_\lambda$, with the relations
\begin{equation}
  \label{eq4}
  q_\lambda=(q_{\lambda 0},{\bf q})\;,\qquad\qquad\qquad
q_{\lambda 0}({\bf q})=\pm\sqrt{M_\lambda^2+{\bf q}^2}
\end{equation}
the normalisation being
\begin{equation}
\label{eq:2o}
<q_{\lambda^\prime}|q_\lambda>=(2\pi)^3\frac{q_{\lambda 
0}}{M_\lambda}\delta_{\lambda\lambda^\prime}\delta({\bf q}-
{\bf q}^\prime)\;.
\end{equation}
The states with $A+1$ particles are denoted  
with the index $\nu$ and is understood to run over all the $A+1$ 
excited states. For them eqs. \eqref{eq:1o} to \eqref{eq:2o} also
hold up to the replacement $\lambda\to\nu$.

Here, in order to be completely covariant we have allowed the intermediate 
states to contain  negative energy solutions too. This case will be 
practically irrelevant in nuclear physics, but  not at all negligible if we 
want to extend this formalism to QCD.

Note also that, just to make a choice, we have assumed for a state with 
baryon number $A$ a boson normalisation ($A$ is assumed to be even). 
Thus consequently
an $A-1$ state must be normalised as a fermion.

In order to make the Lehmann representation for $\Gh$ more explicit we 
introduce in
eq. (\ref{eq1}) the complete set $\sum_\lambda |q_\lambda><q_\lambda|$
(for the $A-1$ system) in the first term of its rhs and
$\sum_\nu |q_\nu><q_\nu|$ in the second one.
We easily find\cite{MoVa-89,CeMoVa-94}
\begin{equation}
  \label{eq1.1}
  \Gh(p,p^\prime;q)=
\sum_\lambda{\varphi_\lambda({\bf p},{\bf q})
\overline \varphi_\lambda({\bf p}^\prime,{\bf q})\over
q_{\lambda 0}({\bf q})-q_0-i\alpha}+\sum_\nu
{\psi_\nu({\bf p}^\prime,{\bf q}^\prime)\overline\psi_\nu({\bf p},
{\bf q}^\prime)\over p_0+p_0^\prime-q_0-q_{\nu 0}({\bf q}^\prime)
+i\alpha}
\end{equation}
where  the $\varphi$ and $\psi$ are defined as
\begin{eqnarray}
  \label{eq1.2}
  \varphi_\lambda({\bf p},{\bf q})&=&\sqrt{M_\lambda\over 
2p_0 q_{\lambda 0}}
<q_\lambda|\psi|p>\\
\label{eq1.3}
\psi_\nu({\bf p},{\bf q}))&=&\sqrt{M_\nu\over 2p_0q_{\nu 0}}
<p|\psi|q_\nu>\;.
\end{eqnarray}
The
equal time commutation relations  imply
\begin{multline}
  \label{eq:bx1}
  \int\frac{d^3y}{2\sqrt{p_0p_0'}}<p'|\left\{\psi^\dagger,\psi({\bf y})\right\}
    |p>e^{-i({\bf p}-{\bf q})\cdot{\bf y}}=\\
    <p'|p>\int\frac{d^3y}{2\sqrt{p_0p_0'}}\delta({\bf y})=
    (2\pi)^3\delta({\bf p}-{\bf p}')~.
\end{multline}
Inserting now a complete set of intermediate states $\sum_{\lambda,\nu}
|q_{\lambda,\nu}>
<q_{\lambda,\nu}|$ in the two terms of the anti-commutator in the lhs and using
\eqref{eq0.5} we get the
a completeness equation in the form
\begin{equation}
  \label{1.4}
  \sum_\lambda\varphi_\lambda({\bf p},{\bf q})
 \varphi^\dagger_\lambda({\bf p}^\prime,{\bf q})
+\sum_\nu\psi_\nu({\bf p}^\prime,{\bf q}^\prime)
\psi^\dagger_\nu({\bf p},
{\bf q}^\prime)=(2\pi)^3\delta({\bf p}-{\bf p}^\prime)~.
\end{equation}

We observe that now the functions $\varphi_\lambda$ and $\psi_\lambda$
play the same role of $\phi_n$ and $\psi_n$ in the eqs. \eqref{eq-1:13}
and \eqref{eq-1:14} of sec. \ref{sec:0}, but now the formalism is Poincar\'e
invariant and further, even if we are only considering the nucleonic
Green's function, all the information about the
dynamics of the system is already embedded in 
$\varphi_\lambda$ and $\psi_\lambda$. 

\begin{figure}[ht]
  \begin{center}
    \centerline{
      \epsfig{file=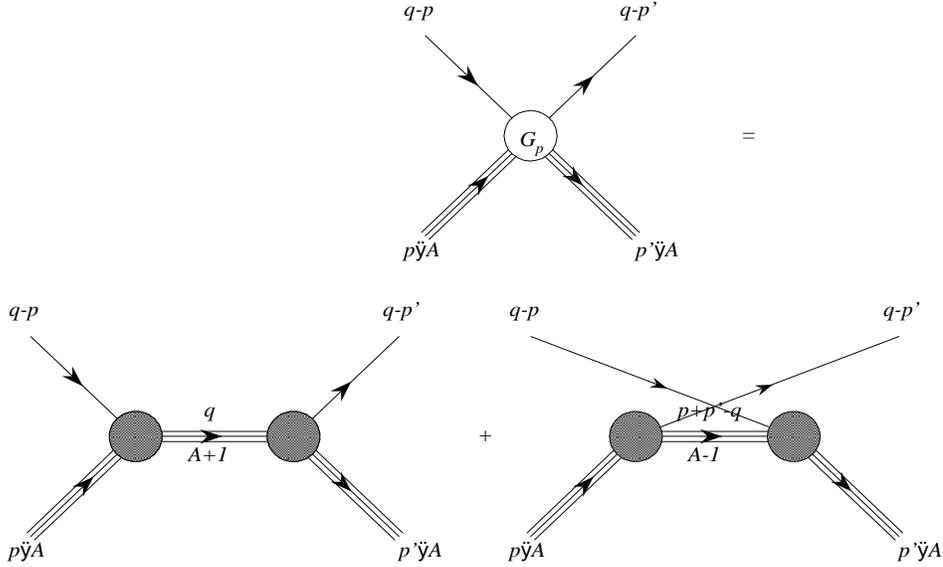,height=8cm,width=14cm}
      }
    \caption{The kinematics for the ``particle'' propagation}
    \label{fig:14}
  \end{center}
\end{figure}

In the same line as above, we can also introduce a ``particle'' 
kinematics: in analogy with $\Gh$ we introduce
\begin{eqnarray}
  \label{eq1:31}
  \Gp(p,p^\prime;q,q_1)&=&\int d^4y\, d^4y^\prime\,G_{pp^\prime}
  (y,y^\prime)
  e^{i(q-p^\prime)y-i(q_1-p)y^\prime}\\
\nonumber
 &=&(2\pi)^4\delta^4(q-q_1)\Gp(p,p^\prime;q)
\end{eqnarray}
whose graphical representation is given in fig. \ref{fig:14}.

By comparison with eq. (\ref{eq0.5.1}) one immediately establishes the link
\begin{equation}
  \label{eq:32}
  \Gp(p,p^\prime;q)=\Gh(p,p^\prime;q')~.\qquad\qquad q'=p+p^\prime-q
\end{equation}

For sake of  simplicity we consider a specific model
of a fermionic field interacting with
scalar bosonic field $\sigma(y)$,
 neglecting the  self-interactions $\sim \sigma^3$ and  $\sigma^4$.
In this simplified scheme (that nevertheless still contains all the 
difficulties  relevant to the fermionic sector) the Hamiltonian of 
the system reads
\begin{equation}
H=\int d^3 
y\overline{\psi}_y\left(-i\bfm{\gamma}\bfm\nabla_y+m+g\sigma_y\right)\psi_y
+H^0_\sigma
\end {equation}
$H^0_\sigma$ being the free Hamiltonian of the $\sigma$
meson).
Using the equation of motion for the field operator $\psi$
\begin{equation}
\gamma^0\left[\psi,H\right]=\bfm{\gamma}\left[\psi,\hat{\bf p}\right]
+(m+\sigma)\psi
\label{eq5}
\end{equation}
we can derive the evolution equation for  $G$, namely
\begin{eqnarray}
  \label{eq1.6}
  \lefteqn{\left(i\gamma\cdot\partial_y-m\right)G_{pp^\prime}(y,y^\prime)=}
  \\&&
  (2\pi)^3\delta({\bf p}-{\bf p}^\prime)\delta^4(y-y^\prime)
  -i\frac{g}{2\sqrt{p_0p_0'}}
  <p^\prime|{\cal T}\left\{\sigma(y)\psi(y)\overline{\psi}(y^\prime)
  \right\}|p>\nonumber
\end{eqnarray}

We can first of all prove that, on general grounds, that 
$G_{pp^\prime}(y,y^\prime)$ can be inverted and consequently 
a mass operator can be defined by means of a perturbation expansion.
The standard proof requires to introduce the generating functional
for connected diagrams and then to perform a L\'egendre transformation on it,
and since it can be found in the usual textbooks \cite{Am-78-B},
is not reported here.
We only remark that the generalisation of the Green's function definition
used in the present paper will only affect the boundary conditions
of the path integral representation of the generating functional,
but not the steps needed to define the mass operator.

We rewrite eq. (\ref{eq0.5.1}) in the form of a Dyson's-like equation 
in Fourier transform as
\begin{eqnarray}
  \label{eq1.6.1}
  \lefteqn{\left\{\gamma\cdot(p-q)-m\right\}\Gh(p,p^\prime;q)=(2\pi)^3
    \delta({\bf p}-{\bf p}^\prime)}
  \\
  &&
  +\intt{{p_1}}\left[\Mh(p,p_1;q)\Gh(p_1,p^\prime;q)
    +\Mp(p_1,p';q')\Gp(p,p_1;q')\right]
    ~,\nonumber
\end{eqnarray}
the mass operators $\Mh$ and $\Mp$ being defined according to the 
``hole'' and ``particle'' channels through
\begin{align}
  \label{eq:ww01}
  \intt{p_1}\Mh(p,p_1;q)\Gh(p_1,p^\prime;q)&=\frac{(2\pi)^3}{2\sqrt{p_0 p_0'}}
  <p'|\bar\varphi\frac{\delta(\hat{\bf p}-{\bf q})}{H-q_0-i\alpha}\sigma
  \varphi|p>\\
  \intt{p_1}\Mp(p_1,p';q')\Gp(p,p_1;q')&=\frac{(2\pi)^3}{2\sqrt{p_0 p_0'}}
  <p'|\varphi\sigma\frac{\delta(\hat{\bf p}-{\bf q}')}{q_0-H+i\alpha}
  \bar\varphi|p>
\end{align}

 In the above 
$p$ and $p'$, as well as $p_1$, are restricted to the mass shell
and $\sigma$ is defined as
\begin{equation}
  \label{eq:w11}
  \sigma=\sigma(y)\bigm|_{y=0}\quad\Rightarrow\quad \sigma(y)=e^{i\hat p y}
  \sigma e^{-i\hat p y}~.
\end{equation}
Further, the index $h$ in $\Mh$ only remind the ``hole'' kinematics chosen
in introducing the Fourier transform. In the configuration space the mass
operator is univoquely determined by the Green's function and embodies 
both particle and hole propagation.

\begin{figure}[th]
  \begin{center}
    \epsfig{file=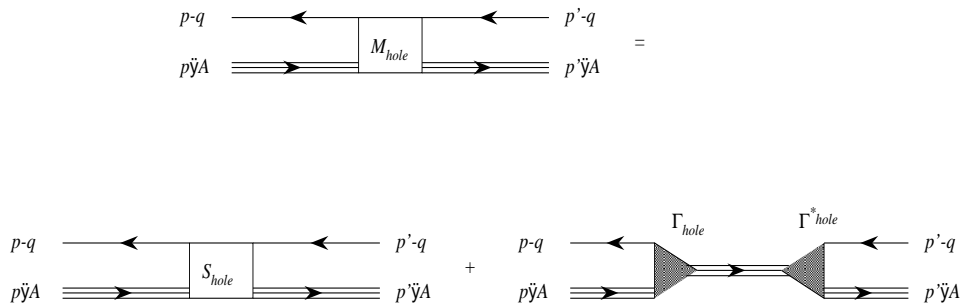,height=8cm,width=13.5cm}
    \caption{The diagrammatic representation of the Dyson's mass operator}
    \label{fig:12}
  \end{center}
\end{figure}

Concerning the structure of the mass operator, the general theory tells us
that it is built by the sum of all 1PI (one-particle-irreducible) diagrams 
defined in sec. \ref{sec:5x} as shown in fig. \ref{fig:12},
and must have the analytical structure
\begin{equation}
  \label{eq:39.1}
  \Mh(p,p',q)=S_{\rm h}(p,p',q)+\sumint \frac{
  {\Gamma_{\rm h}}_n(p,q){\Gamma_{\rm h}}_n^*(p',q)}{{q_{\rm h}}_n({
    \bf q})-q_0-i\alpha}~.
\end{equation}
Here the first term is a smooth (regular) function of $q_0$ while the second
term carries poles in $q_{\rm h}$, living of course in the region 
$q^2>M^2_{A-1}$.
Eq. \eqref{eq1.6.1} also shows that a pole of $\Mh$ corresponds
to a 0 of $\Gh$ and vice versa.

The ``particle'' mass operator has  an analogous expansion but
its poles lie in the region $q^2>M^2_{A+1}$.

The knowledge of the Green's function (or of the mass operator) gives us 
access to many observables, like, for instance, the number of baryons
\begin{equation}
  \label{eq:bx12}
  \begin{split}
  A&=\Tr\int\frac{d^4q}{(2\pi)^4i}\gamma^0\Gh(p,p;q)e^{-iq_0\alpha}
  \end{split}
\end{equation}
and the ground state energy
\begin{equation}
  \label{eq:x4}
  \begin{split}
    p_0&=\frac{<p|H|p>}{<p|p>}\\
    &=-i\Tr\intq{q}\Bigl\{\left[{\mathbf\gamma}\cdot({\bf p}-{\bf q})+m\right]
    \Gh(p,p;q)\\
    &+\intt{p'}\Mh(p,p';q)\Gh(p',p;q)\Bigr\}+\frac{<p|H_0^\sigma|p>}{<p|p>}~.
  \end{split}
\end{equation}

Now we can introduce  the analogous of the ``single particle 
wave functions'' in our relativistic approach, generalising what was described
by eqs. (\ref{eq-1:7},\ref{eq-1:8}). 
This is done by taking eq. (\ref{eq1.6.1}) and integrating it
 over $q_0$ in a small circle containing a pole $q_{\lambda 0}$
(within box normalisation  if needed) and remembering that the 
poles of the Green's function and of the mass operator never coincide. In so 
doing  we immediately find
\begin{eqnarray}
  \nonumber
  \left\{\gamma(p-q_\lambda)-m\right\}\varphi_\lambda({\bf p},{\bf q})
  &=&
  \int{d^3p^\prime\over{2(2\pi)^3\sqrt{p_0p_0^\prime}}}
  \Mh(p,p^\prime;q_\lambda)\varphi_\lambda({\bf p}^\prime
  ,{\bf q})\\
  &=&\sqrt{M_\lambda\over 2p_0q_{\lambda 0}}<q_\lambda|\sigma\psi|p>~,
  \label{eq9}\\
  \left\{\gamma(q_\nu-p)-m\right\}\varphi_\nu({\bf p},{\bf q})
  &=&
  \int{d^3p^\prime\over{2(2\pi)^3\sqrt{p_0p_0^\prime}}}
  \Mp(p^\prime,p;q_\lambda)\varphi_\lambda({\bf p}^\prime
  ,{\bf q})\label{eq:x002}\\
  &=&\sqrt{M_\nu\over 2p_0q_{\nu 0}}<p|\sigma\psi|q_\nu>~.\nonumber
\end{eqnarray}

The above quantities $\varphi$ are {\em not}, of course, wave functions, 
because they feel the presence in the system of antiparticles as well as 
of mesons, but can be looked at as eigenfunction of the system. The case
of uniformly invariant system (free Fermi gas or maybe quark-gluon plasma)
may enable us to make strongly simplifying assumptions. For finite
systems we however can still exploit the idea of a mean field calculation.

\section{The relativistic shell model}
\label{sec:4}

The last equations of the previous section contain the ground idea
to build the relativistic analogue of the shell model.

We first consider the ``hole'' channel and
 rewrite eq. (\ref{eq9}) in the form \cite{Mi-67-B}
\begin{equation}
  \label{eq37:1}
  \left\{\gamma(p-q_\ell(q))-m\right\}\varphi_\ell( p, q)=
\intt{p'}
\Mh(p,p^\prime;q)\varphi_\ell(p^\prime,q)
\end{equation}
with the subtle difference that now $q_0$ is considered a free parameter.
If follows that (\ref{eq37:1}) considered at a given $q_0$ can be
regarded as an eigenvalue equation, the eigenvalue being
$q_{\ell 0}(q)$ that is in general different from the $q_0$ fixed
and contained in $\Mh$. Here notations matter: in fact $\varphi_\ell$
depends upon the 4-vector $q$, chosen by the exterior. We have left the 
dependence upon $p$ instead of ${\bf p}$ to remind the reader that 
$\varphi_\ell$ is a 4-spinor depending, furthermore, upon Lorentz-covariant
quantities like $q^2$ and $p\cdot q$, being understood however that $p_0$
is fixed by the mass shell condition.

Having distinguished between $q_0$ and the eigenvalue $q_{\ell 0}(q)$,
 (\ref{eq37:1}) will have a complete
orthogonal set of eigenfunction, i.e., the $\varphi_\ell$
must obey the properties
\begin{eqnarray}
  \label{eq1:55}
  \intt{p}\varphi_{\ell^\prime}^*(p,q)
  \varphi_\ell( p,q)&=&\delta_{\ell^\prime\ell} c_\ell(q)\\
  \label{eq1:56}
  \sum_\ell \frac{1}{c_\ell(q)}\varphi_\ell(p,q)
  \varphi_\ell^*(p^\prime,q)&=&(2\pi)^3\delta({\bf p}-{\bf p}^\prime)\;,
\end{eqnarray}
$c_\ell(q)$ being a suitable normalisation factor.

Eq. (\ref{eq37:1}) at a fixed and suitably chosen $q_0$ looks like a 
shell model equation having a (non-local) ``shell model potential''
$\Mh$; the functions $\varphi_\ell( p,q)$ are not connected with any observable
quantities, but they are expected, for a reasonable
approximation of $\Mh$ and in a convenient range of
$q_0$ (it means some average of the single particle levels
of a shell model well) to approach the ``single hole'' wave functions
$\varphi_\ell( p, q)$ previously introduced. This is of course 
likely below the Fermi level. 

For the particle channel we rewrite eq. \eqref{eq:x002} as
\begin{equation}
  \label{eq:x003}
  \left\{\gamma(Q_\ell(q)-p)-m\right\}\chi_\ell(p,q)
  =
  \int{d^3p^\prime\over{2(2\pi)^3\sqrt{p_0p_0^\prime}}}
  \Mp(p^\prime,p;q_\lambda)\chi_\ell(p',q)~,
\end{equation}
normalisation and completeness relations being fully analogous the ``hole
wave function'' case.

According to the general form \eqref{eq1.1}
we can now introduce the shell-model-like 
Green's function as
\begin{equation}
  \begin{split}
  \label{eq1:80}
  \gh(p,p^\prime;q)&=
    \sum_\ell \frac{N_\ell}{c_\ell(q)}\frac{
    \varphi_\ell(p,q)\overline{\varphi}_\ell(p^\prime,q)}
  {q_{\ell0}(q)-q_0-i\alpha}\\&
  + \sum_\ell \frac{(1-N_\ell)}{d_\ell(q')}
  \frac{\chi_\ell(p',q')\bar\chi_\ell(p,q')}
  {q'_0-Q_{\ell0}(q')+i\alpha}\Biggm|_{q'=p+p'-q}
  \end{split}
\end{equation}
where $c_\ell$ and $d_\ell$ are suitable normalisation factors and
\begin{equation}
  \label{eq16.1}
  N_\ell=\theta(\ell_F-\ell)~,
\end{equation}
with
\begin{equation}
  \label{eq16.1.1}
  \sum_\ell N_\ell=A~.
\end{equation}
Of course the labels $\ell$'s are ordered increasingly with the 
corresponding energy and $\ell_F$ denotes the highest occupied level
(Fermi level).

The equation the function $\gh(p,p^\prime;q)$ fulfils
is similar to (\ref{eq1.6.1}), provided the $\delta$ function in the r.h.s.
is replaced by
$$\sum_\ell \gamma_0\left\{\frac{N_\ell}{c_\ell(q)}
    \varphi_\ell(p,q)\overline{\varphi}_\ell(p^\prime,q)
  +  \frac{1-N_\ell}{d_\ell(q')}
  \chi_\ell(p',q')\bar\chi_\ell(p,q')\right\}~.$$
The above is of course expected to approximate a $\delta$ as far as the
``shell model'' approximation holds valid.

In the above formalism clearly
 the poles of ${\cal G}_{\rm hole}$ must also be poles
of $\gh$ (the converse is not true in general because some poles of $\Gh$ 
are killed by the projection operator $1-N_\ell$).

We assume that the equations
\begin{eqnarray}
  \label{eq17}
  q_{\ell 0}(q_{\lambda_\ell})&=&q_{\lambda_\ell 0}({\bf q})\\
  Q_{\ell 0}(q_{\nu_\ell})&=&q_{\nu_\ell 0}({\bf q})
\end{eqnarray}
have one or more (maybe infinite) roots for a given ``single particle'' 
quantum number $\ell$. The $q_{\lambda_\ell 0}({\bf q})$ are the eigenvalues of
the equation for the $A-1$ particle state
\begin{equation}
  \label{eq17.1}
  H|q_{\lambda_\ell,\nu_\ell}>=q_{\lambda_\ell\nu_\ell 0}
  |q_{\lambda_\ell,\nu_\ell}>\;.
\end{equation}
The residua of $\Gh$ and of $\gh$ below the 
Fermi level coincide and from eq. (\ref{eq1}) and (\ref{eq1:80}) we obtain
\begin{equation}
  \label{eq18}
  \varphi_{\lambda_\ell}({\bf p},{\bf q})
  \overline{\varphi }_{\lambda_\ell}({\bf p}^\prime,{\bf q})
  =\frac{N_\ell}{ c_\ell(q)}
  \left\{1-\frac{\partial q_{\ell 0}(q)}{
      \partial q_0}\right\}^{-1}\varphi_\ell
  (p,q)\overline{\varphi }_\ell(p^\prime,q)\Biggm|_{q_0=q_{\lambda 0}
  ({\mathbf q})}
\end{equation}

Thus the  above  suggests  to attribute to the $\varphi_{\lambda_\ell}$
the meaning of a generalisation at the relativist level of a
 single particle wave function in a shell model, fully maintaining, 
nevertheless, Lorentz and Poincar\'e invariance. We repeat once more 
that this occurs because we are considering a ``non-diagonal'' single-particle
Green's function where the recoil of the daughter nucleus is accounted for.

Now we make a physical assumption that further narrows us to the 
shell model: we assume that a one particle level $\ell$
is composed by the same sub-levels $\lambda_\ell$ of the exact many-body
problem in such a way that
\begin{eqnarray}
  \label{eq20}
  \lefteqn{\sum_{\lambda_\ell}\intt{q}\left|\varphi_{\lambda_\ell}
      ({\bf p},{\bf q})
    \right|^2}
  \\ \nonumber&&
  =\sum_{\lambda_\ell}\intt{q}\frac{N_\ell}{ c_\ell( q)}
  \left\{1-\frac{\partial q_{\ell 0}(q)}{
      \partial q_0}\right\}^{-1}\left|\varphi_{\ell}( p,q)
  \right|^2\Biggm|_{q_0=q_{\lambda_\ell 0}
    ({\mathbf q})}=N_\ell~'
\end{eqnarray}
This property is not peculiar of a relativistic system,
since the same will happen
in the non-relativistic case.

Now we are ready to make the last step and introduce a phenomenological
``shell model'' potential $V(p,p';\tilde q)$,
with $p$ and $p'$ restricted to the nucleus mass shell
and the 4-vector $\tilde q$ chosen as
\begin{equation}
  \label{eq:aa1}
  \tilde q=(\tilde q_0,{\bf q})~,~~~~~~~~~~~\tilde q_0=\sqrt{{\bf q}^2
    +M^2_{A-1}}~,
\end{equation}
where $M_{A-1}$ is the mass of the daughter nucleus in its ground state
(this choice maintains the analogy with the non-relativistic case:
see, e.g., \cite{Mi-67-B})

Thus the potential $V$ is independent from $q_0$ and
we assume it to be symmetric, i.e., $V(p,p^\prime;{\bf q})
=V(p^\prime,p;{\bf q})$.
From now on we must guess $V(p,p^\prime;{\bf q})$ on phenomenological 
grounds in such a way that
\begin{equation}
  \label{eq:1y}
  \begin{split}
    \left\{\gamma(p-q_\ell(q))-m\right\}\varphi_\ell( p,{\bf q})
    &-\intt{p'}V(p,p';{\bf q})\varphi_\ell(p^\prime,{\bf q})
    \\&=
    \intt{p'}
    \left[\Mh(p,p^\prime;q)-V(p,p';{\bf q})\right]
    \varphi_\ell(p^\prime,q)
  \end{split}
\end{equation}
will be reasonably small.

Once a parameterisation for $V$ has been given we 
can write down the eigenvalue 
equation for the``hole wave functions''
\begin{equation}
  \label{eq41}
  \left\{\gamma\cdot(p-q_\ell({\bf q}))-m\right\}\varphi_\ell
  (p,{\bf q})=\intt{p'}
    V(p,p';{\bf q})\varphi_\ell(p^\prime,{\bf q})
\end{equation}
being 
\begin{equation}
  \label{eq41.1}
  q_\ell=(q_{\ell 0}({\bf q}),{\bf q})~,\qquad
  p=(p_0({\bf p}),{\bf p})~,\qquad
  p'=(p_0'({\bf p'}),{\bf p'})~.
\end{equation}
The index $\ell$ (of course discrete) summarises now
all the quantum numbers pertaining to a given ``one-hole'' state in the
``shell model potential'' $V$.

As an aside, in analogy with the ``hole'' channel, we can introduce an equation
for the ``particle'' channel as
\begin{equation}
  \label{eq41xx}
  \left\{\gamma\cdot(Q_\ell({\bf q})-p)-m\right\}\chi_\ell
  (p,{\bf q})=\intt{p'}
    V(p,p';{\bf q})\chi_\ell(p^\prime,{\bf q})~.
\end{equation}

Up to now covariance has been preserved. The next step is to find a practical
way to solve the ``shell model equation'' \eqref{eq41}. Since it
is not so easy to give an explicit solution of it in a covariant form,
we are forced, in the following, to choose a suitable reference frame
where the equation is particularly simple. Once the solution has been found,
however, we need a procedure to boost it to any reference frame.
This will be our next task.

Thus, coming  to the ``hole'' channel, there are two
 natural choice for the reference frame. One is to assume it as 
the rest frame for the $A-1$ system,
i.e., ${\bf q}=0$:
the eigenvalue equation there becomes
\begin{equation}
  \label{eq:y2}
  \left\{\gamma\cdot k_\ell-m\right\}\varphi(k,0)=
  \intt{k'}V(k,k')\varphi(k',0)
\end{equation}
where
\begin{equation}
  \label{eq:y3}
  k_\ell=(p_0({\bf k})-q_{\ell 0},{\bf k})~,\qquad
  k=(p_0({\bf k}),{\bf k})~,\qquad
  k'=(p_0({\bf k}'),{\bf k}')~.
\end{equation}
Another customary choice  is to assume ${\bf p}=0$
(rest frame for the $A$-nucleons system).
Of course any reference frame can be reached by means of a boost. Thus let
$|0>$ be the ${\bf p}=0$ frame and let $\Lambda$ be the boost from $|0>$
to the state $|k>$ corresponding to ${\bf q}=0$:
$$|k>=\Lambda|0>~.$$
Let us specify in more details the state of the daughter nucleus: we expand
the previous index $\ell$ as
$$\ell\equiv(\lambda,J,M,\pi)$$
where $J$ is the total angular momentum, $M$ its third component and $\pi$
the parity. $\lambda$ will then resume all the other intrinsic (not 
frame-dependent) quantum numbers.
We can now write
\begin{equation}
  \begin{split}
    \label{eq:y11}
    \varphi_{\lambda,J,M,\pi}({\bf k},0)&=\frac{1}{\sqrt{2p_0({\bf k})}}
    <0;\lambda,J,M,\pi|\psi(0)\Lambda|0>
    \\&=
    \frac{1}{\sqrt{2p_0({\bf k})}} <0;\lambda,J,M,\pi|\Lambda\Lambda^{-1}
    \psi(0)\Lambda|0>\\
    &=\frac{1}{\sqrt{2p_0({\bf k})}}S(-{\bf v})
    <-\eta_\lambda{\bf k};\lambda,J,M,\pi|\psi(0)|0>
  \end{split}
\end{equation}
where as usual $S$ is defined through the relation
\begin{equation}
  \label{eq:y12}
  S(-{\bf v})\psi(0)=\Lambda^{-1}\psi(0))\Lambda
\end{equation}
and reads
\begin{equation}
  \label{eq:y13}
  S(-{\bf v})=\sqrt{\frac{p_0({\bf k})+M}{2M}}\left(1-\frac{\gamma^0
      {\boldsymbol \gamma}
      \cdot{\bf k}}{p_0({\bf k})+M}\right)~.
\end{equation}
Of course 
\begin{equation}
  \label{eq:y15}
  {\bf v}=\frac{\bf k}{p_0({\bf k})}
\end{equation}
denotes the velocity of the boost from
 the rest frame of the daughter nucleus
and an extra factor accounting 
for the mass difference between the $A$ and $A-1$ systems is required, namely
\begin{equation}
  \label{eq:y16}
  \eta_\lambda=\frac{M^{A-1}_{\lambda,J,\pi}}{M}~.
\end{equation}
The above entails
\begin{equation}
  \label{eq:y17}
  \Lambda^{-1}|0;\lambda,J,M,\pi>=|-\eta_\lambda{\bf k};\lambda,J,M,\pi>~.
\end{equation}
In the ${\bf p}=0$ frame  the 4-vector $q$ transforms into
\begin{eqnarray}
  \label{eq:y18}
  q'_0\frac{M^{A-1}_{\lambda,J,\pi}}{\sqrt{1-v^2}}&=&
  \frac{M^{A-1}_{\lambda,J,\pi}}{M}p_0({\bf k})=
  \sqrt{(M^{A-1}_{\lambda,J,\pi})^2+\eta_\lambda^2{\bf k}^2}\\
  {\bf q}'&=&-\frac{{\bf v}M^{A-1}_{\lambda,J,\pi}}{\sqrt{1-v^2}}
  =-\eta_\lambda{\bf k}
\end{eqnarray}
and the ``eigenfunction'' $\varphi$ reads
\begin{equation}
  \label{eq:y19}
  \begin{split}
  \varphi_{\lambda,J,M,\pi}(0,{\bf k})&=\sqrt{\frac{M^{A-1}_{\lambda,J,\pi}}
    {2Mq_{\lambda J M\pi;0}({\bf k})}}
  <{\bf k};\lambda,J,M,\pi|\psi(0)|0>\\&=S(-{\bf v}_\lambda)
  \varphi_{\lambda,J,M,\pi}(-{\bf k}/\eta_\lambda,0)
  \end{split}
\end{equation}
with
\begin{equation}
  \label{eq:y20}
  {\bf v}_\lambda=\frac{{\bf k}}{q_{\lambda J M\pi;0}({\bf k})}~.
\end{equation}
This definitions implies
\begin{multline}
  \label{eq:y21}
  \intt{k}\overline{\varphi}_{\lambda,J,M,\pi}(0,{-\bf k})
  \varphi_{\lambda,J,M,\pi}(0,{\bf k})\\=\eta_\lambda^2\intt{k}\overline{
    \varphi}_{\lambda,J,M,\pi}({\bf k},0)\varphi_{\lambda,J,M,\pi}({\bf k},0)
  ~.
\end{multline}

The formalism above has shown that
we can solve the ``shell model'' equation
\eqref{eq41} in the rest frame for the $A-1$ daughter nucleus and then,
using the above kinematics, transfer the solutions to the usual rest
frame of the $A$ nucleus, namely ${\bf p}=0$.
Having established good transformation properties of the solutions, we now 
need to find them in the preferred reference system. Before doing
explicit (model) calculations let us investigate a little what lies 
beyond the shell model.

\section{The  perturbative   expansion}
\label{sec:5x}

The shell model in nuclear physics is usually thought as the 0$^{\rm th}$
order (mean field) of a perturbation expansion. 
Using the eigenfunctions derived from eqs. \eqref{eq41} and \eqref{eq41xx}
we can represent the ``unperturbed'' 
Green's function in the ``hole'' kinematic, in analogy with 
\eqref{eq1:80}, as
\begin{equation}
  \label{eq:a1}
  \begin{split}
    G_{\rm h}^0(p,p^\prime;q)&=
    \sum_\ell \frac{N_\ell}{c_\ell(q)}\frac{
      \varphi_\ell(p,{\bf q})\overline{\varphi}_\ell(p^\prime,{\bf q})}
    {q_{\ell0}({\bf q})-q_0-i\alpha}\\&
    + \sum_\ell \frac{1-N_\ell}{d_\ell(q')}
    \frac{\chi_\ell(p',{\bf q}')\bar\chi_\ell(p,{\bf q}')}
    {q'_0-Q_{\ell0}({\bf q}')+i\alpha}\Biggm|_{q'=p+p'-q}
  \end{split}
\end{equation}

Also, we can introduce a 
 ``shell-model particle Green's function'' by applying the relation 
\eqref{eq1:31} to eq. \eqref{eq:a1}, namely
\begin{equation}
  \label{eq:ax4}
  G^0_{\rm p}(p,p';q)=G^0_{\rm h}(p,p';p+p'-q)~.
\end{equation}

\begin{figure}[h]
  \begin{center}
    \leavevmode
    \epsfig{file=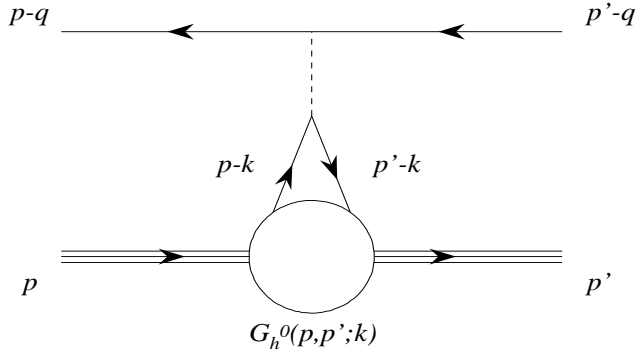,width=10cm,height=6cm}
    \caption{Diagrammatic description of the 
      Hartree contribution to the mass operator}
    \label{fig:44}
  \end{center}
\end{figure}

\begin{figure}[ht]
  \begin{center}
    \leavevmode
    \epsfig{file=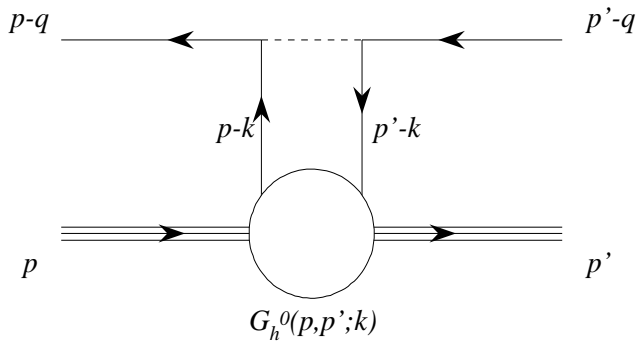,width=10cm,height=6cm}
    \caption{Diagrammatic description of the 
      Fock contribution to the mass operator}
    \label{fig:45}
  \end{center}
\end{figure}

\begin{figure}[thbp]
  \begin{center}
    \centerline{
      \epsfig{file=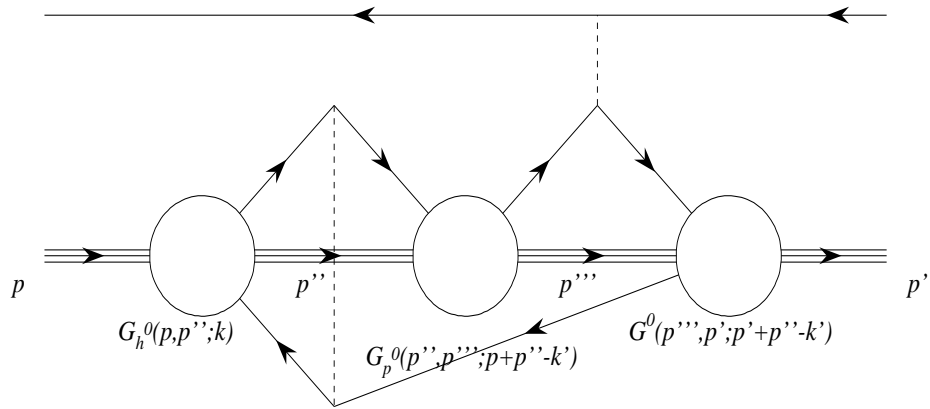,width=14cm,height=8cm}
      }    
    \caption{A typical example of higher order diagram in a perturbative 
      calculation}
    \label{fig:46}
  \end{center}
\end{figure}

The first order in the expansion of the mass operator ${\cal M}$ in terms
of the meson interaction coincides the Hartree-Fock approximation: the first
contribution reads
\begin{equation}
  \begin{split}
    \label{eq:39.2}
    (\Mh)_{\rm Hartree}&=-i\Tr g^2\intq{k}\Gh^0(p,p',k)
    {\cal D}_0(p-p')\\
    &=\Tr\sum_lg^2\intt{k}\overline{\varphi}_\ell(p',{\bf q})
    \varphi_\ell(p,{\bf q})D_0(p-p')
  \end{split}
\end{equation}
where of course
\begin{equation}
  \label{eq:39.3}
  {\cal D}_0(k)=\frac{1}{k^2-m^2_\sigma+i\alpha}
\end{equation}
is the free $\sigma$ propagator, and is displayed in fig. \ref{fig:44},
while
the second term represents, as one can easily convince himself, the Fock 
contribution, namely
\begin{equation}
  \label{eq:39.4}
      (\Mh)_{\rm Fock}=i\Tr g^2\intq{k}\Gh^0(p,p',k)
    {\cal D}_0(k-q)~.
\end{equation}
and is represented in fig. \ref{fig:45},
The other terms have complicated and in practice non manageable 
expressions that involve the detailed structure (i.e., the excited states)
of the target nucleus. We only show diagrammatically a second order 
contribution in fig. \ref{fig:46}.

\section{A simple model}
\label{sec:6}

The above theory looks rather formal. 
Thus 
let us show how it can be implemented
in a practical case. In order to have
manageable formulas we consider a ``shell model potential'' of the 
separable form
\begin{equation}
  \label{eq:y22}
  V({\bf k},{\bf k}^\prime)=-(2\pi)^3\sum_{Jlj}Y_{Jlj}({\bf k})
  Y_{Jlj}^\dagger({\bf k}^\prime)v_J(|{\bf k}|)v_J(|{\bf k}^\prime|)
\end{equation}
where as usual
$$Y_{Jlj}({\bf k})=\sum_s <l,j-s;\frac{1}{2},s|l,\frac{1}{2};J,j>
Y_{lj-s}({\bf k})\chi_s$$
are the generalised spherical harmonics and $v_J$ is some 
function to be chosen in such a way to reproduce the nuclear phenomenology.
Actually we put
\begin{equation}
  \label{eq:y23}
  v_J(|{\bf k}|)=\left(\frac{c}{m^2+{\bf k}^2}\frac{1}{b+e^{k a(A,J)}}
    \right)^\frac{1}{2}
\end{equation}
where $b$ an $c$ are constant and $a(A,J)$ will depend upon the 
atomic number $A$ and, {\em a priori}, upon the total angular momentum $J$.

Now we can solve eq. \eqref{eq:y2} (we assume that,
as established in sec. \ref{sec:4}, once the eigenfunctions
$\varphi_n$ have been found in the frame of reference ${\bf q}=0$
then the above described transformations can provide their expression
in any other frame).

The spinor (not strictly speaking a wave function) solution of  \eqref{eq:y2}
will be labelled by $J$ and $\omega$ and has the form
\begin{equation}
  \label{eq:ax1}
  \varphi_{J,\omega}=
  \begin{pmatrix}
    Y_{Jlj}(\hat{\bf k})F_{J,\omega}(k)\\
    Y_{Jl'j}(\hat{\bf k})G_{J,\omega}(k)
    \end{pmatrix}
\end{equation}
where
$$l=J+\frac{\omega}{2}~,\qquad\qquad\qquad l'=J-\frac{\omega}{2}$$
and
\begin{equation}
  \label{eq:ax2}
  \omega=\begin{cases}
    +1&\text{for~states~with~parity~}(-1)^{J+\frac{1}{2}}\\
    -1&\text{for~states~with~parity~}(-1)^{J-\frac{1}{2}}
  \end{cases}
  ~.
\end{equation}

Using the well known relation
(independent of the parity)
\begin{equation}
  {\boldsymbol\sigma}\cdot{\bf k}Y_{Jlj}(\hat{\bf k})=-kY_{Jl'j}(\hat{\bf k})
\end{equation}
we find
\begin{equation}
  \label{eq:ax7}
  (k\cdot\gamma-m)\varphi_{J,\omega}=
  \begin{pmatrix}
    \left[(k_0-m)F_{J,\omega}(k)+kG_{J,\omega}(k)\right]Y_{Jlj}(\hat{\bf k})
    \\ 
    -\left[kF_{J,\omega}(k)+(k_0+m)G_{J,\omega}(k)\right]Y_{Jl'j}(\hat{\bf k})
  \end{pmatrix}
\end{equation}
and now not only eq. \eqref{eq:y2} is separable thanks to the choice 
of the potential but the large and small components of the spinors are
decoupled and one gets
\begin{subeqnarray}
  \label{eq:ax8}
  (k_0-m)F_{J,\omega}(k)+kG_{J,\omega}(k)&=&
  -v_J(k)X_J[F_{J,\omega}]\\
  -kF_{J,\omega}(k)-(k_0+m)G_{J,\omega}(k)&=&
  -v_J(k)X_J[G_{J,\omega}]
\end{subeqnarray}
where the functional $X_J$ is defined as
\begin{equation}
  \label{eq:ax10}
  X_J[f]=\int\limits_0^\infty p^2dp\,v_J(p)f(p)~.
\end{equation}
Then inverting the above and  
recalling \eqref{eq:y3}, i.e., expliciting the expression
of the eigenvalue $k_0=p_0({\bf k})-q_{\ell 0}$, we get the system
\begin{subeqnarray}
  \label{eq:ax11}
  F_{J,\omega}(k)&=&-\frac{(p_0({\bf k})-q_{\ell 0}
    +m)X_J[F_{J,\omega}]+kX_J[G_{J,\omega}]}
  {(p_0({\bf k})-q_{\ell 0})^2-k^2-m^2}v_J(k)~,\\
  G_{J,\omega}(k)&=&\frac{kX_J[F_{J,\omega}]+(p_0({\bf k})-q_{\ell 0}
    -m)X_J[G_{J,\omega}]}
  {(p_0({\bf k})-q_{\ell 0})^2-k^2-m^2}v_J(k)~.
\end{subeqnarray}
Inserting then these expressions into the definition of the functionals
$X_J[F_{J,\omega}]$ and $X_J[G_{J,\omega}]$
we get for them a homogeneous  system, namely 
\begin{subeqnarray}
  \label{eq:ax13}
  X_J[F_{J,\omega}]&=&B^-_J(q_{\ell 0})X_J[F_{J,\omega}]-
   C_J(q_{\ell 0}) X_J[G_{J,\omega}]~,
  \\
  X_J[G_{J,\omega}]&=&C_J(q_{\ell 0})  X_J[F_{J,\omega}]
  -B^+_J(q_{\ell 0}) X_J[G_{J,\omega}]
\end{subeqnarray}
where we have defined
\begin{eqnarray}
  \label{eq:ax14}
  B^\pm_J(q_{\ell 0})&=&
  \int\limits_0^\infty k^2dk\frac{\left[q_{\ell 0}- p_0({\bf k})\pm m\right]
    v^2_J(k)}
  {(p_0({\bf k})-q_{\ell 0})^2-k^2-m^2}~,\\
  C_J(q_{\ell 0})&=&\int\limits_0^\infty k^3dk\frac{v^2_J(k)}
  {(p_0({\bf k})-q_{\ell 0})^2-k^2-m^2}~.
\end{eqnarray}
The eigenvalue equation generated from the system \eqref{eq:ax13}
is then
\begin{equation}
  \label{eq:ax15}
  R_J(q_{\ell 0})
  =[C_J(q_{\ell 0})]^2-(B^-_J(q_{\ell 0})-1)(B^+_J(q_{\ell 0})+1)=0~.
\end{equation}
Note that the functions $B^\pm_J$ and $C_J$ are real only in the range
$q_{\ell 0}>M_A+m$ or $q_{\ell 0}<M_A-m$, (the former case referring to 
a $A$-nucleus system plus an antinucleon and the latter to a $A$-nucleus  
plus a hole). Here of course $M_A=p_0(0)$ denotes the rest mass of the
$A$-nucleus.

Once the equation is solved in $q_{\ell 0}$
we also get, up to a normalisation constant, the explicit expressions
for the ``wave functions''
\begin{subeqnarray}
  \label{eq:ax16}
  F_{J,\omega}(k)&=&-\frac{p_0({\bf k})-(q_{\ell 0}+m)C_J(q_{\ell 0})
    -k(1-B^-_J(q_{\ell 0}))}{(p_0({\bf k})-q_{\ell 0})^2-k^2-m^2}v^2_J(k)~,\\
  G_{J,\omega}(k)&=&\frac{-(p_0({\bf k})-q_{\ell 0}-m)(1-B^-_J(q_{\ell 0}))
    +kC_J(q_{\ell 0})}{(p_0({\bf k})-q_{\ell 0})^2-k^2-m^2}v^2_J(k)~.
\end{subeqnarray}
Note that, from \eqref{eq:ax15}, the energy levels are degenerate with respect
to $j$ and to the parity and in the notations we can rewrite $q_{\ell 0}$
as $q_{J0}$.

If we further introduce the notation
\begin{equation}
  \label{eq:bx111}
  \varepsilon=M_A-q_{J0}-m
\end{equation}
then the functions $B^\pm$ and $C$ become
\begin{align}
  \label{eq:bx2}
  B^+(\varepsilon)=&-\int\limits_0^\infty k^2dk\frac{\Delta p+\varepsilon}
  {(\Delta p+\varepsilon)(\Delta p+\varepsilon+2m)-k^2}v^2_J(k)\\
   B^-(\varepsilon)=&-\int\limits_0^\infty k^2dk\frac{\Delta p+\varepsilon+2m}
  {(\Delta p+\varepsilon)(\Delta p+\varepsilon+2m)-k^2}v^2_J(k)\\
  \intertext{and}
  C(\varepsilon))=&\int\limits_0^\infty k^2dk\frac{1}
  {(\Delta p+\varepsilon)(\Delta p+\varepsilon+2m)-k^2}v^2_J(k)\\
\end{align}
where we put
\begin{equation}
  \label{eq:bx3}
  \Delta p=p_0({\bf k})-M_A
\end{equation}
to better control the orders-of-magnitude: this last quantity is in fact
expected to be small (say, of the order of $k^2/2 M_A$) unless we look to 
extreme situations, and for bound states $\varepsilon$ is of the order of 
few $MeV$.

To exemplify how the above works,
we have chosen the parameters in \eqref{eq:y23} as
\begin{subeqnarray}
  \label{eq:a22}
  a(A,J)&=&\frac{1}{m}\left[0.7314+0.3274 A^\frac{1}{3}
    -0.0884^\frac{2}{3}+0.0089 A-0.005(2J-1)\right] \nonumber\\~\\
  b&=&0.09\\
  c&=&0.1
\end{subeqnarray}
and we have evaluated the hole energy
for different values of $A$ and $J$. The results (in MeV) are reported
in table \ref{tab:1}.For sake of simplicity we have assumed
\begin{equation}
  \label{eq:a24}
  M_A=A(m+\mu)
\end{equation}
and the chemical potential $\mu$ is chosen as usual as $\mu=-8$ MeV.
\begin{table}[h]
  \begin{center}
    \leavevmode
    \begin{tabular}[h]{||c||c|c|c|c|c||}
      \hline
      \hline
      A&$J=1/2$&$J=3/2$&$J=5/2$&$J=7/2$&$J=9/2$\\
      \hline
      \hline
      12&-9.5&-8&&&\\\hline
      24&-10.5&-9&-8&&\\\hline
      40&-12.5&-10.5&-9&-8&\\\hline
      60&-15&-13&-11.5&-10&-8\\\hline\hline
    \end{tabular}
    \caption{Energy levels of the relativistic ``shell model'' 
      for different $A$ and $J$  }
    \label{tab:1}
  \end{center}
\end{table}

The above example shows how our formalism works. To our knowledge the approach
presented in this paper is beyond the usual relativistic shell calculations,
since the usual ways to afford relativity (see our ref. \cite{Ra-01} and the
many references quoted therein) mainly concern  QHD 
(Quantum-Hadro-Dynamics) inspired models with a space-dependent mass term that
explicitly breaks Poincar\'e invariance. This flaw is obviously not  
obnoxious when heavy nuclei are concerned, it has no future however
when handling a nucleon as a 3-quark system. 

\section{Conclusion and outlook}
\label{sec:11}

In the present paper we have shown how a relativistic theory of the
nucleus can be constructed still preserving the main features of the 
shell model. In our approach in fact a shell model like equation
has been constructed, admittedly in a well defined reference frame,
but we have also built up  all the formalism needed
to boost the results to any other
frame of reference, thus reconstructing Lorentz and Poincar\'e invariance:
this  is a by far non-trivial achievement, since in the traditional 
nuclear physics translational invariance is broken from the very beginning by 
the shell model even in a non-relativistic scheme. Of course the above is
particularly suitable for small systems, since recoil and center-of-mass
motion is fully accounted for. This goes clearly beyond the approaches
based on translationally invariant systems\cite{CeSh-86-B,CeDoMo-97}.

Of course some approximations can be needed in practical calculations, 
and mainly we introduce a ``shell-model potential'' which is thought to 
approximate the mass operator. Again, exactly as described in sec.
\ref{sec:0} we can use the same potential to describe particle and hole 
dynamics, but still the same disease survives, since in principle 
the mass operator in the ``hole'' and ``particle'' kinematics are 
intrinsically
different. Thus we can use the same potential as a starting point, but then
different perturbative expansions are required, as shown in sec. \ref{sec:5x}.

The key issue of the paper is the definition of the ``shell model wave 
function'': we systematically use quotation marks in referring this quantity
because it is not at all a wave function, but is defined, instead, as the
expectation value between physical states (containing any kind of particles,
namely nucleons, mesons and antinucleons) of some field operators. Thus
these quantities, referred to in the above as $\varphi$ and $\chi$, maintain
the formal analogy with the the true nuclear wave functions of the 
non-relativistic shell model, but contain a much more involved dynamics,
since as many mesons and antinucleons as possible are allowed to appear,
and hence accounted for, inside $\varphi$ and $\chi$, the only constraint
being a variation of the baryonic number of $\mp1$.

Thus we can apply the formalism developed so far to any relativistic system,
not necessarily to nuclei, but also (as obvious) to nucleons, where
the ``shell model wave 
function'' (still with quotation marks!) can be regarded as the analogous 
of the quark wave functions in the constituent quark model, not disregarding,
however, the parton content of the constituent quark, which is a composite
object built up on current quarks,antiquaks and gluons. 

Our formalism can also 
be regarded as a theoretical ground for the constituent quark model
and at the same time shows its limitations: in fact, as shown above, we
can derive from the $\varphi$ and $\chi$ the static properties of the
nucleus or of the nucleon, but the response functions (in the nuclear case)
or the $\gamma$ or lepton interaction with a nucleon require a more
detailed study, since the degrees of freedom embodied in the 
``shell model wave function'' require to be explicitly dealt with.
We plan in a successive work to explore the dynamical properties of a 
relativistic complex but finite system.


\end{document}